\documentclass[12pt,twoside]{article}   
\usepackage{epsfig,times}  
\usepackage{color}

\begin{document}
\thispagestyle{empty} 

\markboth{{ \sl \hfill Chapter 1 \hfill \ }}
         {{ \sl \hfill Introduction \hfill \ }}

 \renewcommand{\topfraction}{.99}      
 \renewcommand{\bottomfraction}{.99} 
 \renewcommand{\textfraction}{.0}


\newcommand{\nc}{\newcommand}

\nc{\qI}[1]{\section{{#1}}}
\nc{\qA}[1]{\subsection{{#1}}}
\nc{\qun}[1]{\subsubsection{{#1}}}
\nc{\qa}[1]{\paragraph{{#1}}}

\def\qbu{\hfill \par \hskip 6mm $ \bullet $ \hskip 2mm}
\def\qee#1{\hfill \par \hskip 6mm #1 \hskip 2 mm}

\nc{\qfoot}[1]{\footnote{{#1}}}
\def\qL{\hfill \break}
\def\qpar{\vskip 2mm plus 0.2mm minus 0.2mm}
\def\tvi{\vrule height 12pt depth 5pt width 0pt}
\def\qtvi{\vrule height 2pt depth 5pt width 0pt}
\def\qth{\vrule height 15pt depth 0pt width 0pt}
\def\qtb{\vrule height 0pt depth 5pt width 0pt}

\def\qparr{ \vskip 1.0mm plus 0.2mm minus 0.2mm \hangindent=10mm
\hangafter=1}

\def\qdec#1{\par {\leftskip=2cm {#1} \par}}

\def\qdpt{\partial_t}
\def\qdpx{\partial_x}
\def\qddpt{\partial^{2}_{t^2}}
\def\qddpx{\partial^{2}_{x^2}}
\def\qn#1{\eqno \hbox{(#1)}}
\def\qds{\displaystyle}
\def\qw{\widetilde}
\def\qmax{\mathop{\rm Max}}   
\def\qmin{\mathop{\rm Min}}   

\def\qv{\vskip 0.1mm plus 0.05mm minus 0.05mm}
\def\qhu{\hskip 1mm}
\def\qhv{\hskip 3mm}
\def\qvv{\vskip 0.5mm plus 0.2mm minus 0.2mm}
\def\qhw{\hskip 1.5mm}
\def\qleg#1#2#3{\noindent {\bf \small #1\qhw}{\small #2\qhw}{\it \small #3}\qv }


\color{yellow} 
\hrule height 210mm depth 15mm width 180mm 
\color{black}
\vskip -190mm
\centerline{\bf \Large Fifteen years of econophysics:}
\vskip 4mm
\centerline{\bf \Large worries, hopes and prospects}

\vskip 1.5cm
\centerline{\bf Bertrand M. Roehner $ ^1 $ }
\vskip 4mm
\centerline{\bf Institute for Theoretical and High Energy Physics,
University of Paris 6}

\vskip 5mm
\centerline{\bf 11 April 2010}
\vskip 10mm

\def\qdec#1{\par {\leftskip=1cm {#1} \par}}
\qdec{
\color{blue} \large {\bf Abstract}\quad 
This anniversary paper is an occasion to recall
some of the events that shaped institutional
econophysics. 
But 
in these thoughts about the
evolution of econophysics in the last 15 years we also
express some concerns. Our main worry concerns
the relinquishment of the simplicity requirement. 
Ever since the groundbreaking experiments of Galileo 
some three centuries ago,
the great successes of physicists were largely 
due to the fact that they were able
to {\it decompose complex phenomena into 
simpler ones}. Remember that the first observation of the
effects of an electrical current was made by
Alessandro Volta (1745-1827) on the leg of a frog! 
Clearly, to make sense this observation had to
be broken down into several separate effects.
\qL
Nowadays, with computers being able to 
handle huge amounts of data and to simulate any stochastic
process no matter how complicated, there is no longer any real
need for such a search for simplicity. 
Why should one spend time and
effort trying to break up complicated
phenomena when it is possible to handle them globally? 
On this new road
there are several stumbling blocks, however. Do such 
global mathematical
descriptions lead to a {\it real} understanding? Do they
produce building blocks which can be used elsewhere and thus
make our knowledge and
comprehension to grow in a cumulative way?
Should econophysics also adopt the ``globalized'' perspective
that has been endorsed, developed and spread by the numerous
``Complexity Departments'' 
which sprang up during the last decade?}

\def\qdec#1{\par {\leftskip=2cm {#1} \par}}

\vfill

{\small 1: LPTHE, 4 place Jussieu, F-75005 Paris, France. \qL
\phantom{1:} E-mail: roehner@lpthe.jussieu.fr, Phone: 33 1 44 27 39 16}

\eject

\qI{Past and future}

It is quite by purpose that 
this paper about the 15th anniversary of econophysics is
more focused on the future than on the past. 
Of course, it
will provide an insight about the beginnings of econophysics
but who cares about the past unless it is a springboard 
for the future. The 
main objective of this study is to draw useful lessons
from the past 15 years. Among 
the questions that it will address one can mention
the following points.
\qbu  One of the main objectives of econophysicists was to
apply to social phenomena the methodology that proved so
successful in physics and chemistry. To what extent did they
succeed?
\qbu During the past 15 years, the connection between 
econophysicists and economists did not improve. On the contrary,
the gulf between the two fields became wider.
Can we understand why?
\qbu Historically, physics and chemistry were the fruits of
a search for {\it simplicity}. Indeed, through
their ground breaking experiments
Galileo and Lavoisier were able to lay down basic principles
that provided sound foundations for further progress.
Technical innovation, speculation about natural phenomena,
astronomical observations had existed in various civilizations.
But the small flower of real scientific investigation had
grown only in a few places and time periods. 
To what extent were physicists able to emulate this process 
in the study of social phenomena?
\qbu There has been a massive and worldwide trend in physics 
over the past decade, namely the emergence of 
departments focusing on ``complex systems''.
This term includes a broad range of topics such as
colloids, polymers, sandpiles, traffic jams, neural networks,
colonies of social insects, stock markets, financial derivatives.
The list is endless.
As a matter of fact, almost any real system is complex unless
one has been able to decompose it into simple components. 
For instance,
a pendulum made of a semi-elastic chord and a mass of arbitrary
shape is a complex system in the sense that it can display 
at least half a dozen inter-related effects some of which are
non-linear, e.g. parametric resonance, Foucault effect, 
spatial beat effect, Puiseux effect. This is illustrated in Fig. 1.

%
  \begin{figure}[tb]
    \centerline{\psfig{width=3cm,figure=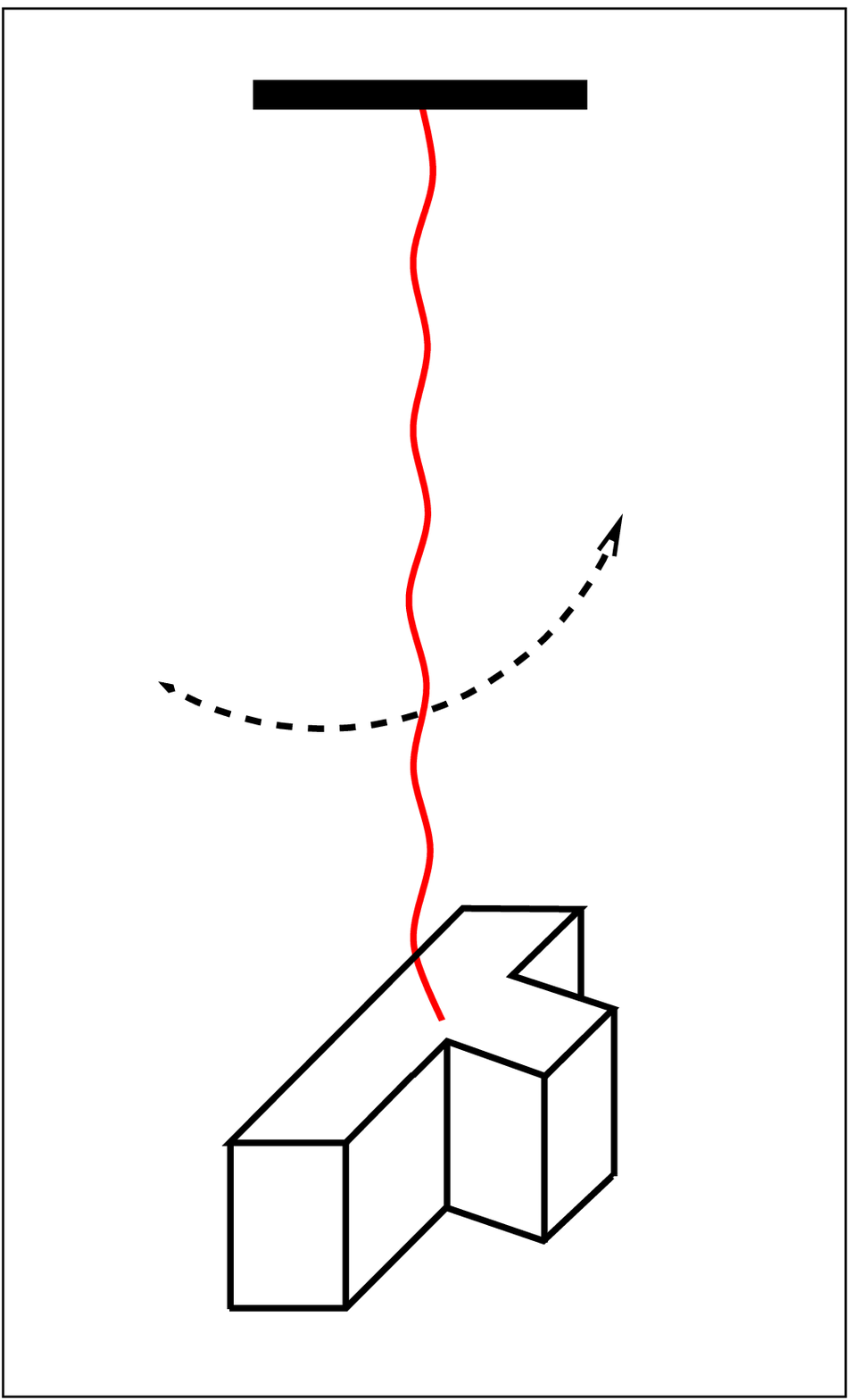}}
\vskip 2mm
\qleg{Fig.\qhu 1\qhv A pendulum seen as a complex system.}
{The pendulum shown here differs from a standard
pendulum only because we have exaggerated some of the 
features which are present in all pendulums. It has a
fairly elastic suspension chord, but of course no
real suspension is completely inelastic. The mass is non-spherical
but no real sphere is completely spherical. This pendulum
shares many of the characteristics of a complex system in the sense
that
it is subject to several inter-connected non-linear effects and
exhibits complicated (yet deterministic)
trajectories. Any 
{\it global} mathematical model (however satisfactory
it may be in terms of adjustment to the observations)
would lead to little
real understanding of the different mechanisms which are at work.
 \qL
Fifty years ago it would have been impossible to study 
this phenomenon globally. Nowadays, the use of computers
have made this possible. Even complicated trajectories
can be recorded and fed into computers and even non-linear
differential equations can be solved numerically and
adjusted to the data through standard curve fitting softwares.
\qL
Through this example we wish to point out that if we give up
trying
to decompose complicated phenomena into simpler sub-modules
we lose much of our ability to get a real insight.
An even more serious consequence is that the models of
complicated phenomena can no longer serve as basic bricks.
Such complex models are just too specific to be transposed elsewhere.
So we lose the ability to build science as a modular structure 
which grows in a cumulative way. 
\qL
In short, even though computers are fantastic tools
in many fields (e.g. engineering or data processing) they
lead physicists to neglect the vital task of simplifying
the phenomena that they wish to understand.}
{}

 \end{figure}

Did this trend benefit econophysics or was it rather disruptive.
Sadly, the prevailing impression is
that this new soft matter physics neglected or threw away
the requirement of simplicity
which for centuries 
had been an essential beacon in physical research.
We will explain why. 
\qpar

Before we begin this review a word of caution is in order.
Obviously, in the limits of a fairly
short paper it was impossible
to do justice to the many books and articles
which contributed to the growth of econophysics. 
Most of the citations will serve to illustrate 
specific events, attempts, trends or ideas that I wish to explain.
This is a fairly personal account. I did not try to
give an objective narrative and apologize in advance to all
the econophysicists whose work, however important, was
not explicitly mentioned.

\qI{The beginnings of institutional econophysics.}

I tried to present an historical account of the beginnings 
of econophysics in an earlier publication 
(Roehner 2002, chapter 2).
In the course of time I came to realize that it has two serious
defects. The first is the fact that it is too Western-centric
in the sense that major events which occurred in China, 
India or Japan were overlooked.
The second is the fact that it does not highlight the
sociological circumstances of the birth of econophysics.
We will come back to these points in a short moment.
\qpar

In spite of its shortcomings the historical account of
2002 nevertheless had one good feature: it made a clear
distinction between pre-institutional econophysics and
institutional econophysics. 
In its institutionalized form, econophysics
came into existence when it became possible to publish
papers on economic or sociological problems in {\it physical}
journals. In the broader sense of ``studies in economics
done by physicists'' econophysics has been in existence 
for a long time. 
Just to mention a few names, Alphonse Qu\'etelet (1796-1874),
L\'eon Walras (1834-1910), Vifredo Pareto (1848-1923) 
or Robert Gibrat (1904-1980) were all physicists or engineers
by education.
\qpar

Now let us come back to the shortcomings.
\qbu In the account of 2002 I gave a list of early conferences 
in econophysics which, quite unfortunately,
overlooked the conferences which were held in India
and in particular the 1995 Kolkata Conference which was 
a satellite meeting of ``Statphys 19'' 
(i.e. the 19th annual conference on
statistical physics) which was held in August 1995 in Xiamen,
(Fujian province, China). It is in the title of a talk given
by Eugene Stanley (published in Physica A as Stanley et al.
(1996)) that the neologism ``econophysics'' was used for the
first time. For a new field it is important to have
a name, otherwise how can one explain to others the kind
of research one is doing.
In subsequent years there have been annual conferences on
econophysics in Kolkata. Entitled ``Econophys-Kolkata I'', the
first was held at the Saha Institute of Nuclear Physics from
15 to 19 March 2005. For a rather informal field such as
econophysics
the regularity of these meetings is quite important and
remarkable.
By bringing about interactions and exchanges, these gatherings
strongly contributed to the development of the new field.
\qbu Any scientific breakthrough has also a sociological
facet. In the first pages of the first chapter of
``Driving forces'' (2007) we describe the sociological facet of
the quantum mechanics revolution which occurred in 
Germany around 1925. At that time, Arthur Compton, 
Paul Dirac, John von Neumann, Linus Pauling, 
Robert Oppenheimer, Edward Teller, Eugene Wigner, none of whom
was German, visited G\"ottingen. It is through their
interaction with 
Born, Einstein, Heisenberg or Schr\"odinger that quantum
mechanics was created and quickly spread to other countries.
Between 1996 and 2000 a similar process was at work
in Boston. Why did it not lead to a revolution of the
same magnitude? The reason is very simple.
\qpar

{\bf Organized versus unorganized data.}\quad
In the decades before 1925 a wealth of experimental
results had been produced and arranged 
in the form of a striking series of empirical laws 
which covered several fields: radioactivity and $ \alpha $
particles,  emission spectra of
atoms, absorption coefficients of X-rays, and so on.
In spite of several phenomenological attempts no comprehensive theory 
had emerged. But the data were there; they were accurate and
already well organized in the form of relationships between
different variables. Nothing of the sort existed in 1995 for
economic or social phenomena. Of course, a lot of
data were available 
but the data were not arranged in the form of clear 
relationships between variables. For instance,
although big amounts of stock prices are available,
these are ``unorganized data'' in the sense that
we do not know what are the main determinants of such prices.
So the time was not really ripe for 
a comprehensive theory.

\qI{An anniversary tribute to Prof. Eugene Stanley}

From the very beginning there have been several 
``schools of thought'' in econophysics, but there can be little doubt
that the Boston school was the most influential. 
Why? Three factors can be mentioned.
\qbu As already mentioned, for several years the
group of Eugene Stanley was a kind of magnet. PhD students, 
postdoctoral students, visitors gathered there. There was a 
steady flow of seminars and at lunch-time everybody 
used to go
to one of the restaurants just across the street. In short, through
discussion and interaction 
this group became the cradle of something new.
\qbu In this group Eugene Stanley held of course a prominent
position, not only through a number of seminal papers (see
Stanley 1995, 1996) but also because of his influence as an
editor of Physica A. In this position and for over 15 years he
nurtured hundreds of econophysical papers.
\qbu As one knows, Sigmund Freud devoted a substantial
amount of time and energy to keeping other 
conceptions than his own
out of the conferences or publications which shaped
the new field. So did Karl Marx. Nothing of the sort
happened in econophysics. There were several schools of thought
at the beginning and 15 years later they are still alive.
In his position Eugene Stanley did not try
to make his own preferences prevail. Thus, in spite of the 
fact that in his own work he was always eager to
establish a close
connection between data and model, completely
theoretical or completely empirical papers were quite
as welcome. \qL
Moreover, in
spite of the fact that econophysics started
with a strong focus on finance, its scope progressively 
broadened. Many papers and books about economic or sociological 
phenomena were published in recent years, for instance on
specific structural regularities in social phenomena
(Wang et al. 2005, Chatterjee et al. 2007, Aoyama et al. 2010) or on the
identification of social interactions (Li et al. 2010,
Zeng et al. 2010). This is certainly a promising trend.
\qpar

So much for history. Let us now turn to an
assessment which can lead to the opening of new roads.

\qI{Theory versus observation}

{\leftskip=2cm My companion prattled away about Cremona
fiddles and the difference between a Stradivarius and an
Amati. 
``You don't seem to give much thought to the matter at hand''
[the Lauriston Garden murder], I said, interrupting
Holmes' musical disquisition.\qL
``No data yet,'' he answered. \qL
``It is a capital mistake to
theorize before you have all the evidence. It biases the
judgment.''
\qpar}
{\leftskip=2cm \hfill  ---Sir Arthur Conan Doyle, 
{\it A Study in Scarlet} (1886) \qpar }

Physicists may not necessarily
agree with the advice given by Sherlock Holmes
in the citation that we put at the beginning.
After all, as we know, Albert Einstein, set up his theory
of gravitation, the General Relativity, purely
from basic principles. At least, this is what we are told.
Two remarks are in order in this respect.
\qbu First, it is not really true that Einstein did not
rely on any data. The effect of the deflection of
a path of light when it is bent by the gravitational field
of the sun was already known in his time
(at least theoretically)
for the simple reason that it is also predicted by Newtonian
mechanics; the two predictions differ by a factor 2, however.
Similarly, the secular change 
in the elliptic orbit of Mercury was well known
from astronomical observations since the mid-19th century.
These observations were
accounted for even by  the earliest versions of the
theory. In other words, it is highly plausible that Einstein had
an eye on such data during the time he was setting up his
theory.
\qbu There is an essential difference between physics at the
beginning of the 20th century and the social sciences at the 
beginning of the 21th century. In 1915, physics was already built on
solid foundations. Any new model or theory had to be consistent
with basic physical principles (e.g. conservation of energy and angular
momentum, laws of reflection or refraction, etc.). On the 
contrary, to my best knowledge in the
social sciences there is not a single 
well-accepted principle, by which I mean one which 
can be tested by observation with a precision better 
than, say,  10\%.
Therefore, it is hardly an exaggeration to say that for
any social phenomenon there are as many models as there are
researches. For instance
at econophysics conferences a wide range of
stock market models have been proposed. All are
able to explain basic characteristics of price
fluctuations but as they do not propose
testable predictions it is impossible to discriminate
between them. The inescapable conclusion is that
this is not the kind of science found in physics.
The ability to discriminate between different
theories is a crucial element in any science.

\qI{Need of a  scientific revolution in the social sciences}

In a recent paper Jean-Philippe Bouchaud (2008) calls
for a scientific revolution in economics. He observes
that economics relies on a set of axioms which, in marked contrast
with the principles of physics, have never been 
really tested by observation. He cites 
the opinion shared by many economists that ``these 
concepts are so strong that they supersede any 
empirical observation''%
\qfoot{Incidentally, 
a few years ago I got exactly the same answer 
to a question that I was naive enough to raise while
attending a course in international macro-economics.}%
.
He observes that during the past decades
the market has been ``deified'' (Nelson 2002).
The fact that actual
achievements of economics have been pathetic 
and disappointing was acknowledged by many prominent 
economists, see for instance
Schumpeter (1933),
Leontief (1982, 1993), Summers (1991), Krugman (2009).
What was the rationale and purpose of the ``market deification''?
Bouchaud points out that 
``the supposed perfect efficacy of a free market stems
from economic work done in the 1950s and 1960s, which with hindsight
looks more like propaganda against communism than plausible science.''
\qpar

However true, Bouchaud's view is perhaps too narrow. 

\qA{The trend toward neoliberal economics}

As a matter of
fact, the target was broader than just communism.   
A vigorous
campaign against Roosevelt's New Deal conceptions
based on social solidarity was orchestrated
by the National Association of Manufacturers even before the 
end of World War II. Some of its earliest steps were the following.
\qbu {\color{blue} Worldwide diffusion of a book by
Friedrich Hayek (1944).} 
Within one year after its 
publication simultaneously in Britain and the
United States, the book was translated  in Spanish, Swedish,
French, Danish. A cartoon version was published in 1945
by {\it Look Magazine}.
\qbu {\color{blue} Creation of the Mont P\'elerin
Society in April 1947.} That this meeting was aimed at being a
public relations event and not just the creation of a 
new scientific society is attested by the fact that among the 39
persons who attended  there were 4 journalists%
\qfoot{The magazines and newspapers which were represented were
the {\it Reader's Digest}, {\it Fortune Magazine}, 
{\it Time and Tide} (a British,
Christian oriented, magazine), and the {\it New York Times.}}
.
\qpar

It makes sense to think that such public relations
paved the way for the economic conceptions that 
became later known as the neoliberal agenda.
\qpar

The attribution of many Nobel prizes in economics 
to this current also played a powerful role.
Among Nobel laureates there have been
5 former presidents of the Mont P\'elerin Society%
\qfoot{F. Hayek (1961, 1974), M. Friedman (1972, 1976),
G. Stigler (1978, 1982), J. Buchanan (1986, 1986), 
G. Becker (1992, 1992); the years within parenthesis
correspond to the end of the terms as president and to
the Nobel prize respectively.}
and
(at least) 3 prominent members of this society%
\qfoot{M. Allais (1988), R. Coase (1991), V. Smith (2002).}%
.
Furthermore,
Erik Lundberg who, for many years, was chairman of
the Nobel Committee was also a distinguished member of
the Mont P\'elerin Society. 
For more details see Roehner (2007, chapter 6). 
.

\qA{Econophysicists seen as dissidents}

If for a moment we admit the thesis set forward by
Robert Nelson (2002) according to which 
the field of economics 
is more a religion and an ideology than a science,
we can perhaps better understand the critical attitude
manifested by
many economists against econophysics
\qpar{Let me mention a personal anecdote in this respect.
In the fall of 1998 I was a visiting scholar at the 
Harvard Department of Economics. 
At some point during a conversation
with my host, Samuel Williamson, I mentioned that
at Boston University, that is to say a few kilometers away,
there was a group of econophysicists led by Eugene Stanley
who was doing path breaking research. My remark did not
attract the slightest attention. 
It is true that some days later a member of this group, namely
Luis Amaral, came to the department to give a talk.
But it was a fairly confidential seminar which was not
attended by any of the main luminaries of the department.}%
\qpar

Why are econophysicists seen as dissenters, dissidents
or competitors who do not play by the rules?
\qbu As econophysicists can publish their research
in physical journals, they are not subject to the rigid
(and fairly arbitrary) control exercised by the
referees of economic journals. 
\qbu Most econophysicists
do not take the main axioms (or beliefs
in Nelson's interpretation) of economics for granted.
Through their insistence on confrontation
with observations and testable predictions, they
clearly try to transform economics into a real science.
Of course, in any religion
priests would feel threatened
by such an undertaking. 
\qbu The mathematisation of economics 
which was undertaken in the 1960s provided a convenient
cover. It contributed to convince the general public 
of the truthfulness of the underlying framework.
``If it is expressed in mathematical formulas and theorems
it must be scientific, isn't it?''.
Because physicists know as much mathematics
as the economists who write in ``Econometrica'' (or in
other journals of mathematical economics), they are
hardly as impressed as the general public. 
They know that any theory, however
elegant and appealing, must be tested.
As far as I know, no Nobel prize in physics has ever been
attributed to a physicist for an
untested theory. In contrast, many
Nobel prizes in economics have been awarded for
untested theoretical researches%
\qfoot{Allais, Arrow, Debreu, Samuelson come to my mind 
in this respect but the actual list is probably 
much longer. In fact, very few Nobel prizes were awarded
for observational research.}%
.
\qpar

The fact that economists do not really
wish to have an honest dialog is exemplified by the
response made to Bouchaud's paper by an economist
of Gothenburg University (Stage 2008). 
He claims
that unemployment rates of the order of 12\% were fairly common
some 50 years ago. As far as I know, 
this is simply not true.
\qpar

\qA{Impact of Bouchaud's paper}

Somewhat sadly, 
Jean-Philippe Bouchaud concludes his assessment by the
following observation. 
``Although numerous physicists have been recruited by financial
institutions over the past few decades, they seem to have forgotten
the methodology of the natural sciences as they absorbed and
regurgitated the existing economic lore.''
This remark clearly addresses what is perhaps the main challenge
to which econophysics is confronted.
\qpar

Bouchaud's paper was written and published in October 2008
that is to say
at the height of the financial panic. Almost two years
have passed.
The panic has been countered by massive injections of
public money, not only to bail out failing financial institutions
but also to buy those toxic assets (e.g. over-the-counter derivatives)
that were spurned by
investment funds. 
Nevertheless, the axioms of the primacy of the market
and of the rationality of agents
have hardly been questioned.
Would such a critical paper have been welcomed
in ``Nature''  prior to the crash and what impact did it left
after the crash?
Between November 2008 and March 2010 three
articles were published in ``Nature'' which discuss or mention
Bouchaud's paper. The first one by Jesper Stage (3 December 2008)
was already mentioned above, the second by
Herv\'e Philippe (7 January 2009)  is a response to Stage; it
weighs our
present economic growth in
the perspective of the long term availability of mineral resources.
Finally in November 2009 there was
a paper by staff writer Mark Buchanan 
which, in 
contrast to Bouchaud's emphasis on the need for empirical 
research and more model-testing, discusses a new
(yet unproven) class of models.

\qI{The big player issue}

{\leftskip=2cm 
``Undue or
unconsidered respect for authority is a prison 
with invisible walls.''
\qpar}
{\leftskip=2cm \hfill  ---Aoyama et al. (2010, Prologue), 
\qpar }

In the economic, social and political worlds there are
so-called {\it big players}%
\qfoot{Also referred to as {\it macro-players} 
in the context of stock markets.(see
Roehner (2006)}. 
By this  expression we mean
big corporations,
big investment funds, influential media corporations
powerful government organizations,
and in a general way all groups which can have a
substantial effect at macro-level whether in the 
economic, political or social sphere.
\qpar

Because many econophysicists come from statistical 
physics they are reluctant to
recognize the role of big players. 
I would suspect that during the past 15 years
the inability or unwillingness of econophysicists 
to recognize the role of big players has hampered their
understanding of many financial, economic or social 
phenomena. The parallel in physics would be to ignore
external forces and treat all systems as being isolated
and solely subject to endogenous interactions. 
Obviously, such an
attitude would lead to serious blunders. That is why this
point deserves careful attention. 
\qpar

Whereas in a crystal
or a liquid  individual molecules cannot 
bring about macroscopic
changes, the physical notion of external factor
nevertheless provides a close parallel to
the concept of macro-player. For instance, a fairly small
voltage at the gate of a field-effect transistor
can control 
a substantial current between source and drain.
An econophysicist who makes the choice of ignoring big players
will be in the same position as a physicist who would try to
explain the movement of electrons in a transistor without
taking into account the effect of the voltage applied 
to its gate.
\qpar

If, just for the purpose of making this argument
somewhat more concrete,
one wishes to get an idea of the power of 
government organizations 
here is a little experiment that can easily be performed 
on the Internet.
\qpar

{\color{blue} Distribution of films: a case-study.}\quad
It turns out that a recent movie entitled
``Ghost writer''  by Roman Polanski which
portrays the hidden influence 
(whether hypothetical or real is irrelevant in the
present argument) of the US State Department
on British policy had a very limited distribution
in spite of being an excellent thriller as attested by the
fact that it received the Silver Bear award
at the Berlin International Film Festival in February 2010.
\qL
According 
to the Mojo Box Office website 
it was screened in only 5 countries, namely
Austria, France, Germany, 
Poland and the United States. The fact that in the US it was
shown in 
only 14 theaters during the opening week (compared with over
2,000 theaters
for a standard Hollywood movie
\qfoot{As a matter of comparison, 
the film ``Green zone'' on the
occupation of Iraq seen from the American side which
was released almost the same
week as ``Ghost writer'' opened in 3,002 theaters 
across the United States.}%
) 
suggests that very few 
American people were in fact able to watch it.
The movie got a good reception and wide audience 
in all European countries where it was shown.
Yet, quite amazingly, up to now (4 April 2010)
it has not been released in Britain. I was told by an English
colleague that the release may occur after the general
elections of 6 May 2010. This will be an interesting test.
\qpar

As an additional observation it can be added that the novel
by Robert Harris on which the movie is based was so far
translated into 10 languages%
\qfoot{Chinese (2009), Danish, Finnish, Flemish, 
French (2007), German (2007), 
Italian (2007), Japanese (2009), Spanish, Swedish.}%
.
Usually, according to the publisher, 
Robert Harris' best-sellers are translated in up to
30 languages.
\qpar

Needless to say, this is a fairly elusive issue because the 
US State 
Department will never admit that it took any action
to curb the screening of the movie. On the contrary, such
allegations would be dismissed as being pure fabrication.
Nevertheless, careful scientists and historians will
not necessarily take such denials at face value.
\qpar

{\color{blue} Role of investment funds in overnight stock
crashes.}\quad 
A similar problem occurs when the
price of a stock falls by some 20\% overnight, as happened
for instance for Ebay shares in late January 2005.
In such cases it is
very difficult, if not altogether impossible, to 
know who among of the big share holders has dumped millions
of shares. Yet, it is very unlikely that changes which occur
in such a short time can be the result
of the spread of a panic among small stock holders. 
Through the reports issued by
the Security and Exchange Commission it is possible to
know the percentages held by major stock holders%
\qfoot{Such data are also listed on the Yahoo finance website.}%
.
Unfortunately, such reports are only published every quarter
which makes any precise identification
impossible.
\qpar

{\color{blue} A case illustrating the power of the media.}\quad
As a last example of the role of big players, one can mention
the influence of mass media. This is illustrated in
Fig. 2 by the specific case of media campaigns on the 
occasion of car recalls by automobile
companies. 

%
\begin{figure}[tb]
    \centerline{\psfig{width=16.5cm,figure=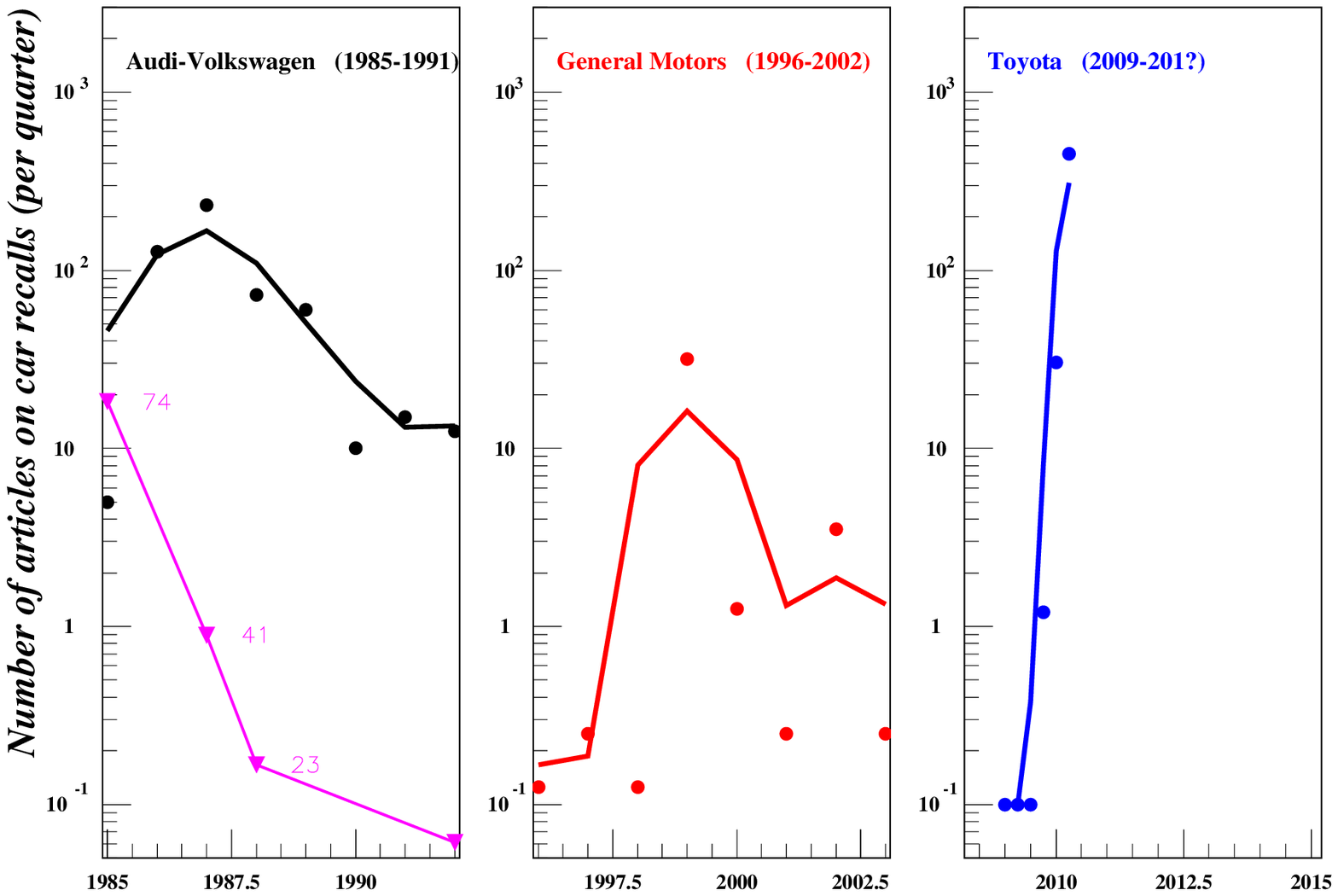}}
\vskip -5mm
\qleg{Fig.\qhu 2\qhv Illustration of the power of 
US media, seen as a particular instance of
the power of big players.}
{The graph documents the action of the US press
(newspapers and magazine) in three recalls of cars.
The interesting point is that in terms of number
of articles published
these actions differed
by several orders of magnitude.\qL
In the Audi case the real start of the campaign
came with a CBS program on 
``60 Minutes'' (November 1986). In particular, it
showed the tragic case of a mother whose Audi car had
run over her son, allegedly
because of a sudden and unexpected acceleration.
However, in March 1989 a ruling of the
``National Highway Traffic Administration''
stated that there was nothing 
wrong technically with Audi cars
and attributed this (and similar) accidents
to mistakes made by drivers. Nevertheless, due to
the media campaign, the sales of Audi cars in the
United States fell from 74,000 in 1985 to 23,000 in
1989 and 12,000 in 1992 (magenta curve with numbers
expressed in thousands of cars sold).
\qL
In the Toyota case, US sales have begun to plunge 
in early 2010 but
it is still too early to say how much and for how long 
they will eventually fall.\qL
From a methodological perspective the main point is that
it would be absurd and pointless
to forget the exogenous forces and study
the spread of this ``rumor''
as being a spontaneous process like the diffusion
of molecules in a liquid.\qL}
{Sources: The information about the ruling of the
National Highway Traffic Administration and about
Audi sales is from the
New York Times (11 March 1989 and 18 April 1993).
The frequency data about the numbers of articles
were obtained from a newspaper data base (Factiva)
through key-word searches. We used the following key-words: 
``Audi+recall+acceleration'', 
``Chevrolet+recall+tank'', ``Toyota+recall+acceleration''.
The numbers have been normalized to take into
account the fact that the coverage provided by the data base
became more systematic in the course of time.}

 \end{figure}

Here again the management of
TV channels or newspapers 
would deny any allegation concerning the very existence of such 
campaigns. They would say that media outlets simply reflected
the concerns of the public. Yet, how can one understand
that the American public should feel more concerned 
about a few fatal cases (allegedly) due to
unexpected accelerations than
about the 13,580 people who died in vehicle fires between
1981 and 2001 of which a substantial proportion
was caused by faulty gas tank design and gave rise to
law suits against General Motors%
\qfoot{The source of this figure is a 2002 report of
the ``National Fire Protection Association'' cited in
``When motorists get burned'', Factiva database,
1 April 2002.}%
?  
\qpar

In short, ignoring the intervention of big players
makes us unable to account for many important 
social effects. In physical experiments
it is possible to minimize exogenous forces.  In
social phenomena unless such forces are identified
and estimated, it makes little sense to study 
endogenous effects.

\qI{An agenda for the next decades}

In order for econophysics to develop in a cumulative
way, the following tasks may be of importance.

\qA{Measuring social interactions}
Many economic phenomena in fact are rather social effects.
For instance, 
when a fashion or a panic spreads like wildfire this phenomenon
has little to do with micro- or macro-economics but is rather
conditioned by the nature and strength of the social links 
between agents. Therefore we must learn how to
measure these social interactions.

\qA{User friendly databases}
Sherlock Holmes had an encyclopedic knowledge of the annals
of crime in European countries as well as in America.
For each case he was able to cite a dozen {\it similar}
cases that had occurred in the last century. In
``The Valley of Fear'' (1915)
he observes that ``if you have all the details of a thousand
cases at your finger ends, it is odd if you can't unravel the thousand
and first.''
We should do the same.
For instance, let us create a data base
for real estate price crashes or credit crunches (often 
the two occur together)%
\qfoot{According to a review published
on the economic history website eh.net, a
recent book by Reinhart and Rogoff (2009) 
seems to fulfill such a program, at least partially. 
I have not yet seen the book itself however.}%
.
Apart from price series, such a data base
should also include sale volumes, information 
about changes in tax regulation or interest rates.
Apart from data on 
residential real estate it should also include data on office
and commercial real estate.  \qL
Other phenomena for which it would be of interest to 
establish comparative data bases are mentioned in Roehner 
(1997, p. 20-23).

\qA{Taking into account the role of big players}
Develop our knowledge of the actions of big players.
Thanks to the Internet, this has now become possible
in many cases as illustrated above.

\qA{A nice research field for econophysics: demography}

Demographical problems are certainly ``simpler''
\qfoot{It would be possible to give to this notion
a more precise definition in terms of
diversity of agents and interactions,
range of time scales, frequency of endogenous and exogenous
shocks, and so on.}
than stock markets
but  at first sight it might seem that there are
no challenging questions in demographical phenomena.
This is not true. 
As a matter of fact,
we still do not understand the determinants of many effects.
For instance, 
predictions of fertility rates one decade ahead in time
made by American demographers of the Department of
the Census
in 1950, 1955, 1962, 1972 consistently
turned out to be wrong
by wide margins between 50\% and 100\% (see
Roehner 2004b, Fig. 9.2 p. 328).
Another example of mystery is
the huge death rate 
(several times higher than the death rate of married
males of same age) of young widowers.
The younger the widowers, 
the larger the difference (Roehner 2007, p. 194).
\qpar

For physicists demography has two nice features: 
(i) There are a lot of accurate data 
(ii) Basic demographic phenomena are a good field for comparative
analysis across different countries and time periods.

\qA{Model testing}
For a majority of the physicists with whom I have been able 
to discuss it seems clear that it is the testing procedure
of models which must be improved if we wish to converge toward
some basic underlying principles. The key-question is how
such tests should be conducted.
\qpar

First of all, it can be observed that the very existence of
econophysics led to a notable improvement in this respect.
Many predictions have been proposed by econophysicists
during the past 15 years especially for real estate prices and
stock market prices, see for instance:
Roehner (2001, p. 176),  Sornette (2003, p. 373), 
Roehner (2004a, p. 115), Roehner (2006, p. 179), 
Richmond (2007, p. 286).
\qpar

On the contrary,
economists do not usually offer predictions.
We do not include in this assessment the short-term
predictions made by macro-economists working for
governmental statistical agencies. Of course, if nothing new happens
the econometric equations which were able to describe the
economy during the last semester may also hold for the next.
But this is just an exercise in parameter 
adjustment and extrapolation, it does not reveal an
understanding of specific mechanisms. 
\qpar

In physics the most important class of predictions are not 
predictions in the course of time. Of course, astronomers
can predict eclipses or planet positions
a few decades ahead in time
but such predictions are successful
only because the level of noise is low enough. 
This is illustrated by the case of meteorology.
Although the 
phenomena which take place in the atmosphere are well-known
physical effects, weather forecasting remains difficult.
This is not because we do not understand 
the underlying phenomena, but because there is a higher level
of noise. Similarly, in most social systems there is
a high level of noise; so, even independently of  
inherently chaotic effects, forecasts will be difficult and
unreliable.
\qpar

In physics
the most common form of predictions consist in
what can be called ``structural'' predictions. 
For instance, once
one has understood the mechanism of the Foucault pendulum
and once this understanding has been tested in a few places,
it becomes possible to predict what will
be the behavior of a Foucault pendulum anywhere else whether
in Rome (i.e. far from the Equator), in Thanjavur (Tamil Nadu, i.e.
near the Equator) or in Sydney (i.e. beyond the Equator
in the southern hemisphere).
\qpar

At this point one should devote some attention
to what is really meant by the expression ``Foucault pendulum''.
This brings us to a point which may be very important
for the future of econophysics.

\qI{The Pareto filtration method in physics and economics}

Clearly, a Foucault pendulum is a pendulum which displays
the Foucault effect. This, however, is a very small effect which
can be over-ridden by several others. 
\qbu For instance if the
mass of the pendulum is not perfectly spherical
(and it never is, of course) this will produce a rotation 
of the pendulum's oscillation plane which has nothing to do with
the Foucault effect. 
\qbu If the length of the pendulum is too small
the rotation of its oscillation plane 
will be strongly affected by the so-called
Puiseux effect%
\qfoot{The formula discovered by Puiseux (1820-1883) says that:
$ f_p=[3/(16\pi^2)](S/L^2)f $ where $ f $ is the frequency of the 
pendulum, $ L  $ its length, 
$ S $ the area of its semi-elliptic trajectory and
$ f_p $ the frequency with which this semi-ellipse
revolves around its center.}%
.
\qpar

How do experimental physicists get rid of such 
unwelcome phenomena? The answer is very simple.\qL
{\color{blue} \it
Physicists build their experimental devices in  a way which will
filter out spurious effects. \qL
Should we not try to do the same for economic and social phenomena?}
\qpar

What does this mean in practice? 
Before discussing this method
for social phenomena, it must be emphasized that 
even in the natural sciences this is a challenging task.
This can be illustrated by the following example.\qL
In 1907 an Americal medical doctor, Dr.
Duncan MacDougall, weighted six patients about to die
from tuberculosis just before their death and after their death.
He found that on average they were 21g lighter after their
death%
\qfoot{MacDougall's paper was published in the
``Journal of the American Society for Psychical Research'', May 1907
(the full text is available on the Internet) and his experiments 
were described
in the New York Times of 11 March 1907 (p. 5).}%
.
Naturally, before one can admit MacDougall's 
interpretation according
to which the 21g might represent the weight of the 
departing soul, one must
eliminate all other possible factors such as transpiration or
a change in the
volume of air contained in the body. After a careful discussion
MacDougall concludes that these factors cannot account for the
21g difference.  This is of course obvious for the air because
it would represent a volume of 17 liters.
In addition MacDougall conducted similar
experiments on dogs and found no difference in weight.
\qpar

This may seem a weird example but it clearly shows that the
task is not an easy one. Often the only way to get a real
answer is to perform additional experiments. For some unknown
reason Dr. MacDougall did not repeat his experiment on humans,
so we may never know the real answer.
\qpar

Next we illustrate the filtration procedure through
an example in economics.
Suppose we study {\it speculative} real estate price peaks.
As a first, step we would collect as many 
housing price series as possible that
have sharp price increases, say a doubling over
a period of 4 years. Some of these series will
correspond to the phenomenon that we wish to study
while others may be due to a different effect.
For instance, once
a new industry is established in a country town
this will lead to its expansion and also probably to
a substantial rise in rents and house prices. 
Basically this process is different from
a speculative process (athough there can be a
speculative component).
Thus, if we wish to focus on a well-defined phenomenon
such outliers must be discarded. Vilfredo Pareto (1919) was
one of the first sociologists
to advocate such a filtration procedure and to use it
in a systematic way (Pareto 1919).
\qpar

Every time I tried to convince economists that it was
important to use a sound filtration method
they argued that this was a completely arbitrary 
procedure. They did not realize that the success of physics 
over the last three centuries
largely relied on such filtration methodologies. They
allowed physicists to study one
effect at a time instead of dealing with
multifactorial phenomena in which everything is mixed up. 
\qpar

Of course, I am ready to recognize that this procedure is
more difficult to implement for social phenomena than it is in
physics. 
This is mainly because in order to change the circumstances
of an observation one must find real episodes in which
the new conditions are realized. 
Even though this may not be an easy road, 
I do not know of any other way to get
to the {\it core of a phenomenon}%
\qfoot{Pareto calls this core the ``residue''. This word
refers to the operation of filtration in chemistry.
The residue is the (useful) solid component which remains at the
bottom of the filter.}%
.
Pareto used this methodology with success. This should
be a strong encouragement.
The Coriolis effect, the Foucault effect, the Puiseux effect
(and indeed all physical laws) are basic building blocks
which, once identified and well understood, can 
be used to develop physics in a modular and cumulative way.
Personally, I am convinced that similarly there are
basic building blocks in social phenomena.
\qpar

During the past 15 years ``Complex systems'' groups
were set up in many physics departments. 
If the acknowledged purpose of a research group is
to study {\it complex} systems, why indeed should
the researchers bother to devote time and effort to make
them simpler?
One should perhaps not generalize from a few cases,
but I often
got the impression that such complex systems are 
investigated by using sophisticated devices 
which produce large amounts of data but that the
later do not lead to a clear understanding of the mechanism
at work. 
\qpar

As one knows, the Santa Fe Institute was one of the
first institutions
to start this trend. Its motto, which
is printed on the sweaters sold in the lobby of 
the institute, says:
``From simplicity to complexity''. When I stayed
there for a few months in 2002 I tried to explain 
why I would be more happy with a motto saying:
``From complexity to simplicity''. I am not
sure that my remark was well understood.
\qpar

I must confess that sometimes during the past
10 years, especially
while attending ``Complex Networks'' conferences%
\qfoot{One should remember 
that at such conferences about 75\% of the lectures are
theoretical papers without any connection with real
observations.}%
, 
I have been feeling like a swimmer who
tries to swim upstream
but is inexorably dragged in the opposite direction
by the strong current.

\vskip 15mm

{\bf \large References}

\qparr
Aoyama (H.), Fujiwara (Y.), Ikeda (Y.), Iyetomi (H.), 
Souma (W.) 2010: Econophysics and companies.
Statistical life and death in complex business networks.
Cambridge University Press, Cambridge (to appear).

\qparr
Bouchaud (J.-P.) 2008: Economics needs a scientific revolution.
Nature 455, 1181, 30 October.

\qparr
Buchanan (M.) 2009: Waiting for the maths.
Nature Physics  5, 776-776.

\qparr
Chatterjee (A.), Sinha (S.), Chakrabarti (B. K.) 2007:
Economic inequality: Is it natural? Current Science 92, 1383-1389.

\qparr
Harris (R.) 2007: The ghost.
Simon and Schuster, New York.

\qparr
Hayek (F.A.) 1944: The road to serfdom. 
University of Chicago Press, Chicago. 

\qparr
Krugman (P.) 2009: How did economists get it so wrong?
New York Times, 6 September. 

\qparr
Leontief (W.) 1982: Academic economics. Science 9, 17, 104-107.

\qparr
Leontief (W.) 1993: Can economics be reconstructed as an empirical
science? American Journal of Agricultural Economics, October 2-5.

\qparr
Li (X.), Li (M.), Hua (Y.), Di (Z.), Fan (Y.) 2010:
Detecting community structure from coherent oscillation 
of excitable systems
Physica A 389, 1, 164-170.

\qparr
Mantegna (R.N.) 1991: L\'evy walks and enhanced diffusion in 
Milan stock exchange.
Physica A, 179, 2, 232-242 (December).\qL
{\it \color{blue} [This article marks the beginning
of institutional  econophysics in the sense that, along 
with an article by Hideki Takayasu et al. (see below),
it was accepted in
Physica A by editor Eugene Stanley with the deliberate
purpose of starting a new field of application for
statistical mechanics.]}

\qparr
Nelson (R.) 2002: Economics as religion. From Samuelson to
Chicago and beyond.
Pennsylvania State University Press, University Park.

\qparr
Pareto (V.) 1919: Trait\'e de sociologie g\'en\'erale.
Payot, Paris. \qL
This is the French translation of 
``Trattato di sociologia generale'' (1916). \qL
The work
was translated into English under the title ``The mind and
society'' (1935) which reflects better the real purpose
of the author.

\qparr
Qu\'etelet (A.) 1835: Physique sociale ou essai sur le
d\'evelopement des facult\'es de l'homme. 
Bachelier, Paris\qL
The book was republished in 1869 but it seems that it
had never been translated into English. 
A possible translation of the title would read:
``Physics of social phenomena. An essay on human
development.''

\qparr
Qu\'etelet (A.) 1848: Du système social et des lois 
qui le régissent. 
Guillaumin, Paris.\qL
Translation of the title: ``On the laws of the social system''.

\qparr
Reinhart (C.M.), Rogoff (K.S.) 2009: This time is different:
eight centuries of financial folly. 
Princeton University Press, Princeton.

\qparr
Richmond (P.) 2007: 
A roof over your head; house price peaks in the UK and Ireland.
Physica A, 375, 1, 281-287.

\qpar
{\it \color{blue} [The four references which follow
are examples
of early attempts in what became later known as econophysics.
As a matter of fact, there have been other similar attempts
made by physicists;
see for instance
Montroll and Badger (1974) or Weidlich and Haag (1983).
In contrast with the institutional phase of econophysics
which came later,
these papers were not published in physical journals.]}

\qparr
Roehner (B.M.) 1982: Order transmission efficiency in
large hierarchical organizations.
International Journal of Systems Science, 13, 5, 531-546.

\qparr
Roehner (B.M.) 1984: Macroeconomic regularities in the
growth of nations: an empirical inquiry.
International Journal of Systems Science, 15, 9, 917-936.

\qparr
Roehner (B.M.) 1989: An empirical study of 
price correlations. The decrease of price correlation with
distance and the concept of correlation length.
Environment and Planning A, 21, 289-298.

\qparr
Roehner (B.M.) 1996: The role of transportation cost
in the economics of commodity markets.
American Journal of Agricultural Economics 78, 339-353 (May).

\qparr
Roehner (B.M.) 1997: The comparative way in economics: a 
reappraisal. 
Economie Appliqu\'ee 50, 4, 7-32.

\qparr
Roehner (B.M.) 2004a: Patterns of speculation in real estate
and stocks. 
{\it in} Hideki Takaysu, editor: The application of econophysics. 
Proceedings of the Second Nikkei 
Econophysics Conference, Tokyo 12-14 November 2002. 

\qparr
Roehner (B.M.) 2004b: Coh\'esion sociale [Social cohesion].
Odile Jacob, Paris [In French].

\qparr
Roehner (B.M.) 2006: Macro-players in stock markets.
{\it in} Hideki Takaysu, editor: Practical 
fruits of econophysics.
Proceedings of the Third Nikkei 
Econophysics Conference, Tokyo 9-11 November 2004. 

\qparr
Roehner (B.M.) 2007: Driving forces in physical, biological,
and socio-economic phenomena.
A network science investigation of social bonds and 
interactions.
Cambridge University Press, Cambridge. 

\qparr
Roehner (B.M.) 2006: Real estate price peaks.
A comparative overview.
Evolutionary and Institutional Economics Review, 2, 2, 167-182.

\qparr
Schumpeter (J.) 1933: The common sense in econometrics.
Econometrica, 1, 5-12.

\qparr
Sornette (D.) 2003: Why stock markets crash. 
Critical events in complex financial systems. 
Princeton University Press, Princeton.

\qparr
Stage (J.) 2008: Speaking up for economic-sciences modelling.
Nature 456, 570-570.

\qparr
Stanley (H.E.), Buldyrev (S.V.), Goldberger (A.L.), 
Goldberger (Z.D.),
Havlin (S.), Mantegna  (R.N.), Ossadnik (S.M.), Peng (C.-K.),
Simons  (M.) 1994: 
Statistical mechanics in biology: how ubiquitous are 
long-range correlations?
Physica A, 205, 1-3, 214-253.

\qparr
Stanley( H. E.),  Afanasyev (V.),  Amaral (L. A. N.), 
 Buldyrev (S. V.),
 Goldberger(A. L.), Havlin (S.), Leschhorn  (H.), 
 Maass (P.), Mantegna (R. N.),
 Peng (C.-K.), Prince (P. A.), 
 Salinger (M. A.), Stanley (M. H. R.),
 Viswanathan (G. M.) 1996:
Anomalous fluctuations in the dynamics of complex systems: from DNA
and physiology to econophysics
Physica A, 224, 1-2, 302-321 (February).\qL
{\it \color{blue} [The paper contains the following concluding expectation.
``It may be that as we study economies less regulated
or more regulated we will find as rich a phenomenology 
as was discovered to
describe the
various universality classes in critical phenomena. 
And it may be that theoretical models
to explain our empirical findings will be forthcoming.'']}

\qpar
{\it \color{blue} [In subsequent years  (as is attested by
their numbers of citations)
the two following articles became two reference-pillars of the new
field of econophysics.]}

\qparr
Stanley (M.H.R.), Amaral (L.A.N.), Buldyrev (S.V.), Halvin (S.),
Leschhorn (H.), Maass (P.), Salinger (M.A.), Stanley (H.E.) 1996:
Scaling behavior in the growth of companies. Nature 379, 804-806,

\qparr
Mantegna (R.N.), Stanley (H.E.) 1996: Turbulence and financial
markets. Nature 383, 587-588.

\qparr
Summers (L.H.) 1991: The scientific illusion in empirical
macroeconomics.
Scandinavian Journal of Economics, 93, 2, 129-148.\qL
{\it \color{blue} [This is a rare instance of a paper written by a renowned
economist who is highly critical of the actual achievements
of econometrics.]}

\qparr
Takayasu (H.), Miura (H.), Hirabayashi (T.), Hamada (K.) 1992:
Statistical properties of deterministic threshold elements: the case
of market price.
Physica A, 184, 1-2, 127-134 (June).\qL
{\it \color{blue}
[This paper was one of the first econophysical papers ever published
in Physica. It
contains the following prophetic statement:
``Statistical physics and economics have only little connection 
so far but
we think their potential intersection should be very large because in
both fields
the main theme is to clarify the behavior of macroscopic variables in systems
where a great number of microscopic agents are interacting''.
Of course, such a statement also applies to sociology or
political science.]}

\qparr
Wang (D.), Menghui (L.), Di (Z.) 2005:
True reason for Zipf's law in language
Physica A 358, 2-4, 545-550.

\qparr
Weidlich (W.), Haag (G.) 1983: Concepts and
models of a quantitative sociology: the dynamics of interacting
populations. Springer, Berlin.

\qparr
Zeng (A.), Hu (Y.), Di (Z.) 2010: Unevenness of loop location
in complex networks. To appear in Physical Review E.

\end{document}